\pdfoutput=1
\documentclass[12pt]{article}

\usepackage{graphicx,psfrag,epsf,color}
\usepackage{amsmath,amssymb,amsfonts}
\usepackage{hyperref}
\usepackage{array}
\usepackage{bm} 
\usepackage{cite}
\usepackage{hyperref}
\usepackage{multirow}
\usepackage{rotating}
\usepackage{slashed}
\usepackage{textcomp}
\newcounter{MBQ}

\newcounter{KUQ}

\newcommand{\lessim}{\mbox{\raisebox{-3pt}{$\stackrel{<}{\sim}$}}}

\newcommand{\be}{\begin{equation}}
\newcommand{\ee}{\end{equation}}
\newcommand{\bea}{\begin{eqnarray}}
\newcommand{\eea}{\end{eqnarray}}
\newcommand{\bi}{\begin{itemize}}
\newcommand{\ei}{\end{itemize}}
\newcommand{\ben}{\begin{enumerate}}
\newcommand{\een}{\end{enumerate}}
\newcommand{\bt}{\begin{tabular}}
\newcommand{\et}{\end{tabular}}

\newcommand{\mchi}{m_\chi}
\newcommand{\mW}{m_W}
\newcommand{\mZ}{m_Z}

\begin{document}
\allowdisplaybreaks

\begin{titlepage}

\begin{flushright}
{\small
TUM-HEP-1390/22\\
March 2, 2022
}
\end{flushright}

\vskip1cm
\begin{center}
{\Large \bf 
Matching resummed endpoint and continuum\\[0.2cm] 
$\gamma$-ray spectra from dark-matter annihilation}\\[0.2cm]
\end{center}

\vspace{0.45cm}
\begin{center}
{\sc M.~Beneke$^{a}$, K.~Urban$^{a}$,} and  {\sc M.~Vollmann$^{b}$}\\[6mm]
{\it ${}^a$Physik Department T31,\\
James-Franck-Stra\ss e~1, 
Technische Universit\"at M\"unchen,\\
D--85748 Garching, Germany}
\\[0.3cm]
{\it ${}^b$Institut f\"ur Theoretische Physik,\\
Auf der Morgenstelle 14, 
Eberhard Karls Universit\"at T\"ubingen,\\
72076 T\"ubingen, Germany}
\\[0.3cm]
\end{center}

\vspace{0.55cm}
\begin{abstract}
\vskip0.2cm\noindent
For the minimal wino and Higgsino benchmark models 
we provide accurate energy spectra of high-energy photons 
from TeV scale dark-matter annihilation $\chi\chi\to \gamma+X$ 
by merging electroweak Sudakov resummation near maximal 
energy with the electroweak parton-shower PPPC4DM, and 
accounting for the Sommerfeld effect. Electroweak resummation 
significantly changes the shape of the photon-energy spectrum 
in the wide range $E_\gamma \sim (0.6\ldots 1)\, m_\chi$ and 
hence the form of the so-called ``line-signal''. 
\end{abstract}
\end{titlepage}

\section{Introduction}
\label{sec:introduction}

The properties of dark matter (DM) beyond its gravitational
interaction is one of the biggest open questions of particle physics and cosmology
today. Even though not detected so far, weakly interacting massive particles
(WIMPs) remain among the most promising candidates for particle DM, 
especially if connected to the electroweak scale as, e.g., in
supersymmetric models \cite{Arkani-Hamed:2006wnf}, or due to their 
minimal model assumptions \cite{Cirelli:2005uq,Cirelli:2007xd}. 
The two obvious benchmark scenarios for such TeV electroweak WIMPs are the pure 
wino and Higgsino. Their thermal masses are in the $(1-3)$ TeV region
\cite{Beneke:2014hja,Beneke:2020vff}, and a discovery or exclusion is out of
reach for current collider experiments \cite{Canepa:2020ntc}. It is also not clear if future possible
collider experiments will exclude or discover the wino and Higgsino
conclusively \cite{EuropeanStrategyforParticlePhysicsPreparatoryGroup:2019qin}. The situation is
comparably grim for nuclear scattering given cross sections near or
within the coherent neutrino background in direct detection experiments
\cite{Chen:2019gtm}. However, the indirect detection of a TeV WIMP DM
annihilation signal in cosmic rays offers a way to challenge these
models with current and upcoming experiments
\cite{Abdallah:2018qtu,Acharyya:2020sbj}.

One of the most sensitive indirect detection probes is the $\gamma$-ray line
signal. However, only including the line-signal $\chi \chi \to \gamma \gamma +
\frac{1}{2} \gamma Z$ into the analysis is too naive. The  precise prediction 
of the expected DM signal is complicated by several effects. First,
since the DM is electrically neutral, the annihilation into photons is only
possible through loop processes
\cite{Bergstrom:1997fh,Bern:1997ng,Ullio:1997ke}, and more importantly, via
mixing with a slightly heavier charged multiplet partner through the Sommerfeld
effect \cite{Hisano:2003ec,Hisano:2004ds,Beneke:2014gja}. The latter enhances
the cross section by up to several orders of magnitude and is itself subject to
large electroweak (EW) corrections
\cite{Beneke:2019qaa,Beneke:2020vff,Urban:2021cdu}. Second, the large ratio
of DM mass to the EW gauge boson masses together with the semi-inclusive
($\gamma+X$) nature of detecting a photon at earth \cite{Baumgart:2017nsr,Beneke:2018ssm} leads to large
Sudakov double logarithms $\ln^2 (4\mchi^2 / \mW^2)$ that require
resummation
\cite{Baumgart:2014saa,Baumgart:2014vma,Bauer:2014ula,Ovanesyan:2014fwa,Ovanesyan:2016vkk}.
On top, the finite energy resolution, which in the TeV regime for typical
Cherenkov telescopes is of order several percent of the DM mass $\mchi$, induces
further large logarithms of the energy resolution vs. DM mass and/or
electroweak scale
\cite{Baumgart:2017nsr,Baumgart:2018yed,Beneke:2018ssm,Beneke:2019gtg,Beneke:2019vhz}.

The anticipated large energy bins make it imperative to consider the full
spectrum beyond the nominal endpoint at which Sudakov resummation is most
important. The purpose of this letter is to demonstrate how the endpoint 
spectra can be utilized and merged with dedicated parton-shower calculations 
(in this paper PPPC4DM \cite{Cirelli:2010xx}) away from the endpoint 
to obtain realistic predcitions for the ``photon line-signal'' with 
state-of-the-art theoretical precision. To this end, we
investigate the logarithmic structure of resummed and parton-shower 
spectra and demonstrate
which logarithms are correctly captured by either calculation beyond their
naive region of validity. With these insights at hand, we devise a merging procedure, thereby providing differential spectra
close and away from the endpoint. The logarithmic analysis is not
restricted to wino and Higgsino DM but applies to all TeV EW WIMPs, in
particular, the minimal DM candidates \cite{Cirelli:2005uq,Cirelli:2007xd} or
even the full MSSM \cite{Beneke:inprep}. Additionally, 
we provide the code \texttt{DM$\gamma$Spec} ancillary to this paper that supplies ready-to-use 
$\gamma$-spectra, which 
include Sommerfeld and Sudakov resummation to next-to-leading order
(NLO), respectively NLL' (NLL + NLO), for wino and Higgsino DM, merged with
PPPC4DM spectra away from the endpoint.
Details on the code (and download information) and the 
merging procedure is provided in the appendices to the main text.


\section{Endpoint spectrum resummation}
\label{sec:summaryEFT}

The main focus of this paper lies on the exemplary pure wino and Higgsino DM models, that produce the observed relic density with 
a thermal mass of $\mchi = 2.842\,{\rm TeV}$ \cite{Beneke:2020vff} and
$\mchi = 1.1\,{\rm TeV}$, respectively. The former is an SU(2)$_L$ triplet of
Majorana fermions with a mass difference of $\delta \mchi = 164.1\,{\rm MeV}$
between the charged and neutral states of the multiplet \cite{Yamada:2009ve,Ibe:2012sx}. The Higgsino is an SU(2)$_L$ doublet of
Dirac fermions with hypercharge, that splits into two neutral Majorana fermions
$\chi_1^0$ and $\chi_2^0$ with a mass difference $\delta m_N \geq 150\,{\rm
keV}$, that we fix to $\delta m_N = 20 \,{\rm MeV}$ for figures in this paper,
and a charged component of the multiplet, that is $\delta m_\chi \approx 355 \,{\rm
MeV}$ heavier \cite{Thomas:1998wy}. In both models, the detection of a line signal
becomes possible, as the neutral DM particles can convert into 
a pair of charged virtual states before annihilation via
the exchange of EW gauge bosons, commonly known as Sommerfeld effect
\cite{Hisano:2003ec,Hisano:2004ds,Cirelli:2007xd}. The EW 
potentials are
known to NLO for the Higgsino \cite{Urban:2021cdu} and wino
\cite{Beneke:2019qaa,Beneke:2020vff}, and are included to this accuracy 
in this work.

In indirect detection of the photon signal, the observable is
not the literal line signal, i.e., annihilation to $2\gamma \gamma +  \gamma Z$, 
but rather $\chi^0 \chi^0 \to \gamma + X$, where the unobserved final state $X$ 
is  kinematically constrained to be jet-like by the finite energy resolution of the detector \cite{Beneke:2018ssm}. Additionally, for
TeV DM masses, the hierarchy between DM and EW scale masses $\mchi \gg
\mW$ induces large Sudakov double logarithms. The small quantities associated
with these large logarithms are
\begin{align}
\epsilon = \frac{\mW}{2 \mchi}\,, \quad \quad 1-x = 1 - \frac{E_\gamma}{\mchi}
\label{eq:ratiologs}
\end{align}
with $E_\gamma$ the energy of the detected photon. The resummation of these
large logarithms 
is achieved in the framework of soft-collinear effective theory (SCET)
\cite{Bauer:2000yr,Bauer:2001yt,Beneke:2002ph}. Depending on the relative
scaling of $1-x$ and $\epsilon$, different treatments are needed to describe 
the differential spectrum near the endpoint $x = 1$. Resummed 
results are available for:
\begin{itemize}
\item line signal only ($\gamma + X = 2 \gamma \gamma  +\gamma Z$) --- (wino to
    NLL' \cite{Ovanesyan:2014fwa,Ovanesyan:2016vkk})
\item narrow resolution $1-x \sim \epsilon^2$ --- (wino
    \cite{Beneke:2018ssm,Beneke:2019gtg} and Higgsino \cite{Beneke:2019vhz} to
    NLL')
\item intermediate resolution $1-x \sim \epsilon$ --- (wino
    \cite{Beneke:2019gtg} and Higgsino \cite{Beneke:2019vhz} to NLL')
\item wide resolution $1 \gg 1-x \gg \epsilon$ --- (wino to NLL
    \cite{Baumgart:2017nsr,Baumgart:2018yed})
\end{itemize}
where LL, NLL refer to the resummation of the leading, respectively,
next-to-leading logarithms. The NLL' approximation, in addition, includes the
full one-loop corrections. The line signal-only case is contained in the 
narrow resolution calculation by taking the limit $x \to 1$ and decoupling 
the light fermions.
All results can be cast into the form 
\begin{eqnarray}
\frac{d (\sigma v)}{d E_\gamma} &=& 2 \sum_{I,J} S_{IJ} \Gamma_{IJ}^{\rm res} \, ,
\label{eq:resummed_diff}
\\[-0.65cm]
&&\nonumber
\end{eqnarray}
where the sum over $I,J$ runs over the two-particle states that mix with
$\chi^0 \chi^0$ via Sommerfeld enhancement, $S_{IJ}$ encapsulates the
Sommerfeld enhancement and $\Gamma_{IJ}^{\rm res}$ is determined in the
respective effective field-theory (EFT) computation.

\begin{figure}[t]
\centering
\hskip-1cm
\includegraphics[width=0.7\textwidth]{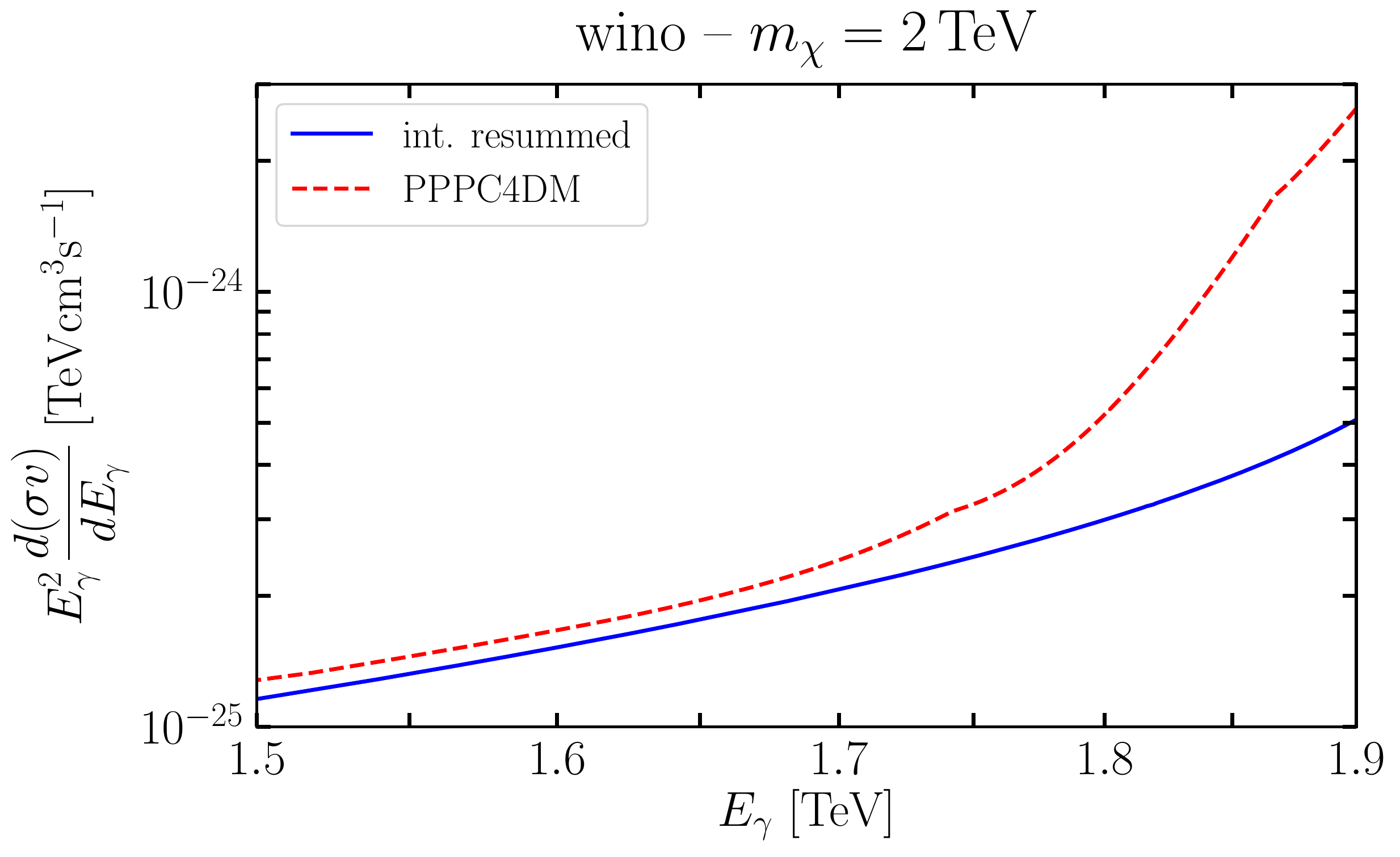}
\caption{Intermediate resolution resummed (blue/solid) and PPPC4DM 
(red/dashed) gamma spectrum for wino annihilation with mass $m_\chi=2 \,{\rm
TeV}$.
}
\label{fig:diff_comp}
\end{figure}

Given the typical energy resolution of Cherenkov telescopes of several percent of DM mass (e.g., \cite{Acharyya:2020sbj}), the
bulk of the endpoint spectrum probed lies in the intermediate resolution
regime. Here, we are concerned with obtaining a full spectrum beyond the
endpoint, and hence consider the merging of the intermediate resolution
logarithms to the parton-shower calculation provided in PPPC4DM
\cite{Cirelli:2010xx}. An example of the spectra using only EFT (intermediate
resolution) or only PPPC4DM is shown for a $2 \, {\rm TeV}$ wino in
Fig.~\ref{fig:diff_comp}, which demonstrates that away from the endpoint, the
Sudakov resummed EFT and PPPC4DM calculations
converge.\footnote{Towards the endpoint in Fig.~\ref{fig:diff_comp}, the
resummed spectrum is suppressed with respect to PPPC4DM. Part of this
suppression is due to the resummed Sudakov logarithms. In addition, PPPC4DM
smears the tree-level delta-distribution at the endpoint, which further
enhances the difference between the two curves in Fig.~\ref{fig:diff_comp}.
For a comparison, including the behaviour at the absolute endpoint of both
calculations, see Fig.~\ref{fig:gaussian} and the accompanying discussion in
Sec.~\ref{sec:conclusion}.} We define
\begin{align}
\left. \frac{d(\sigma v)}{dx} \right|_{\rm PPPC4DM} = 2 \sum_{IJ} S_{IJ}
    \left(\hat{\Gamma}^{WW}_{IJ} \frac{dN_{WW}}{dx}+ \hat{\Gamma}^{ZZ}_{IJ}
    \frac{dN_{ZZ}}{dx} + \hat{\Gamma}^{\gamma Z}_{IJ} \frac{dN_{\gamma Z}}{dx}+
\hat{\Gamma}^{\gamma \gamma}_{IJ} \frac{dN_{\gamma \gamma}}{dx}\right),
\label{eq:dNdx_def}
\end{align}
where $dN_{AB}/dx$ denote the splitting
functions of $AB$ into $\gamma +X$, and  $\hat{\Gamma}^{AB}_{IJ} = (\sigma v)^{IJ \to AB}_{\rm tree}$ are the tree-level annihilation matrices. The inclusion 
of the Sommerfeld enhancement mandates the consideration of
off-diagonal terms such as $\hat{\Gamma}^{AB}_{(00)(+-)}$, which mix the different
neutral two-particle states. Since the wino and Higgsino are
pure multiplets, they only annihilate into EW gauge bosons at tree-level.
The Higgs boson and SM fermions enter only at loop-level or
via the EW evolution. 


\section{Logarithmic structure of the result}
\label{sec:logs}

To allow for a comparison of the logarithmic structure between the EFT
calculations and PPPC4DM, we extend the definition of $dN/dx$ to the resummed
case. Strictly speaking, $dN/dx$ is only useful within the collinear
approximation, which only includes collinear splittings of the 
two-gauge boson final-state of tree-level annihilation. The
resummed calculations on the other hand, also includes model-dependent initial-state radiation (ISR), which cannot be associated with a particular
tree-level annihilation matrix. For the purpose of comparison, we nevertheless
define
\begin{align}
\frac{d N^{IJ}_{WW}}{dx} \equiv 
\mchi \, \frac{\Gamma^{\rm res}_{IJ}}{\hat{\Gamma}_{IJ,\rm tree}^{WW}}
\quad \quad \quad (x<1)
\label{eq:dNdx}
\end{align}
in a slight abuse of notation. We restrict to $x<1$ to exclude virtual corrections, which are proportional to $\delta(1-x)$, from the comparison, 
as we are interested in the differential terms in $x$.

We investigate the leading terms in the $\hat{\alpha}_2$ expansion 
of the two calculations for the wino model.\footnote{Similar 
considerations hold for the Higgsino model, but the formulas become 
lengthier due to the additional $\chi_1^0 \chi_1^0 \to ZZ$
tree-level annihilation term.} The Sommerfeld term
becomes $S_{IJ} = \delta_{I(00)} \delta_{J(00)}$, and the 
right-hand side of~\eqref{eq:resummed_diff} reduces to $\chi^0
\chi^0 \to W^+ W^- \gamma$ in the 
tree approximation. We should mention that this process is 
numerically subdominant (for TeV scale DM masses) 
relative to formally higher-order 
but Sommerfeld-enhanced loop processes such as $\chi^0 \chi^0 \to \chi^+ \chi^- \to W^+ W^- \gamma$ which
is of $\mathcal{O}(\hat{\alpha}_2^5 \mchi^2 / \mW^2)$. 
Nevertheless, we begin
our investigation with $\chi^0 \chi^0 \to W^+ W^- \gamma$, as on top of the
endpoint-resummed \cite{Beneke:2019gtg} and the collinear-approximation result~\cite{Cirelli:2010xx,Ciafaloni:2010ti}, the full fixed-order
computation is also available~\cite{Bergstrom:2005ss}.

The intermediate-resolution endpoint-resummed expression, expanded 
back in $\hat{\alpha}_2$ to lowest non-vanishing order, 
results in the approximation 
\begin{align}
\left.\frac{dN_{WW}}{dx} \right|_{\chi^0 \chi^0\to WW \gamma}^{\rm int.} = \frac{2 \alpha_{\rm em}}{\pi} \left[\frac{1}{1-x} \ln \left(1+\frac{(1-x)^2}{\epsilon^2}\right) - \frac{1-x}{\epsilon^2 + (1-x)^2}\right]\,.
\label{eq:00WWgam}
\end{align}
This approximation implicitly assumes $1 -x \sim
\epsilon \ll 1$. Similarly, we can extract the
corresponding splitting function for PPPC4DM following \cite{Ciafaloni:2010ti},
which employs the collinear approximation, and obtain
\begin{align}
\left. \frac{dN_{WW}}{dx}\right|_{\rm PPPC 4 DM} = \frac{2 \alpha_{\rm em}}{\pi} \left[\frac{x}{1-x} \ln \frac{(1-x)^2}{\epsilon^2} - \left(\frac{1-x}{x} + x (1-x)\right) \ln \epsilon^2\right] \, .
\label{eq:PPPCWW}
\end{align}
The collinear approximation is valid if 
$\epsilon \ll 1-x =\mathcal{O}(1)$.
Finally, we consider the fixed-order unresummed computation
\cite{Ciafaloni:2010ti,Bergstrom:2005ss}, expanded in $\epsilon$ 
and the mass difference between charged and neutral DM, which yields
\begin{eqnarray}
\left.\frac{dN_{WW}}{dx}\right|_{\chi^0 \chi^0 \to WW \gamma}^{\rm f. o.} &=& \frac{2 \alpha_{\rm em}}{\pi} \left[\frac{(1-x+x^2)^2}{(1-x)x} \ln \frac{1}{\epsilon^2} \right. \nonumber \\
&&\hspace{-1.5cm}- \frac{(4-12x+19x^2-22x^3+20x^4-10x^5+2x^6)}{(2-x)^2(1-x)x} \nonumber \\
&&\hspace{-1.5cm}\left.+ \frac{8-24x+42x^2 - 37 x^3+16x^4 - 3 x^5}{(2-x)^3 (1-x)x} \ln (1-x) \right] \, .
\label{eq:NLO00WWgam}
\end{eqnarray}
The fixed-order computation is valid whenever no large logarithms 
appear, i.e. for $1-x \gg \epsilon$ and $x \gg \epsilon$. Hence, we expect the
collinear approximation to match the fixed-order result, for $1-x \gg
\epsilon$.  Eq.~\eqref{eq:NLO00WWgam} 
captures model-dependent terms, which are not
part of the collinear approximation \eqref{eq:PPPCWW}.

\begin{figure}[t]
\centering
\includegraphics[width=0.7\textwidth]{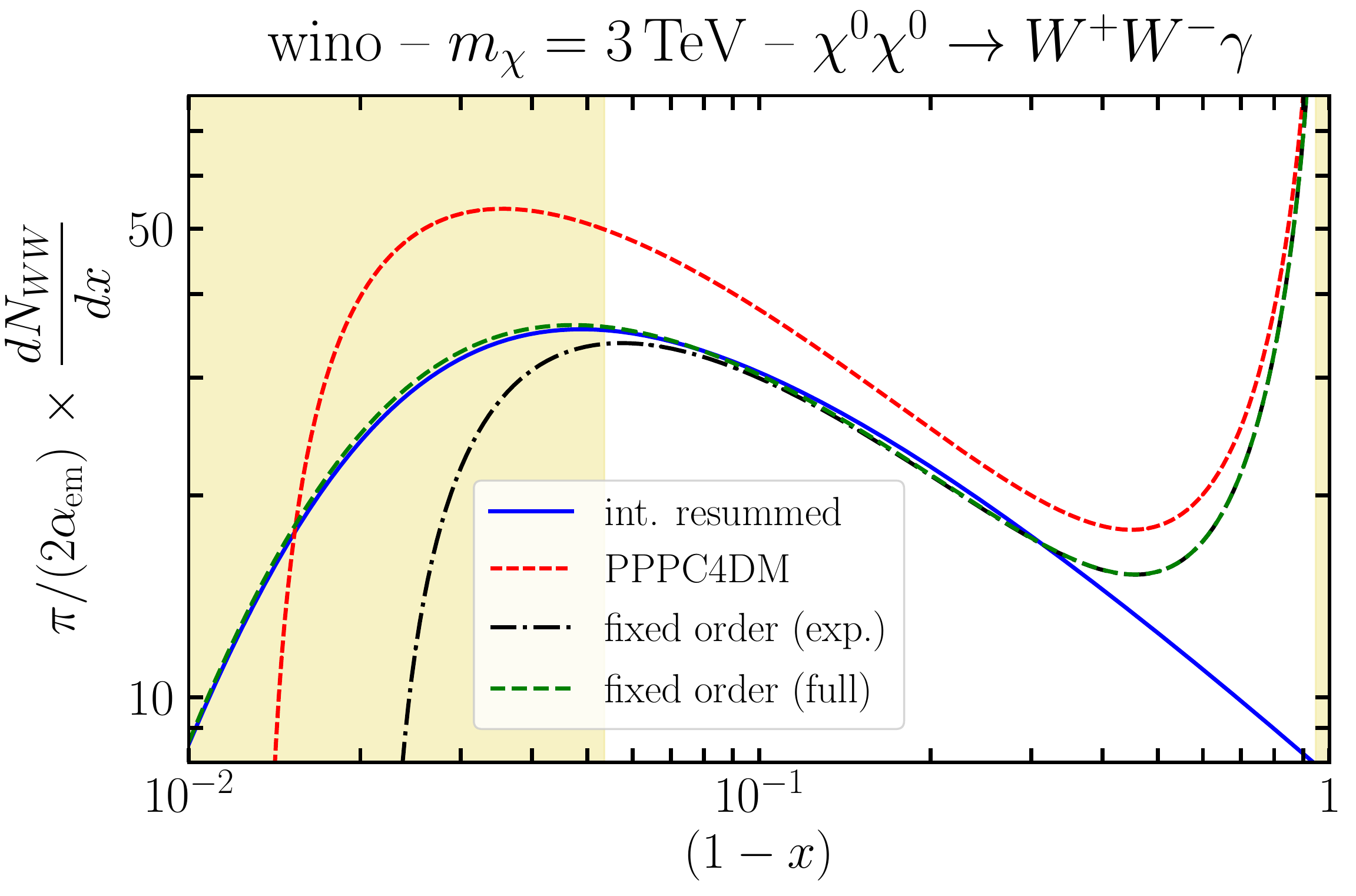}
\caption{Photon energy spectrum from $\chi^0 \chi^0 \to W^+ W^- \gamma$ for a 3 TeV wino ($\epsilon \approx 0.0134$) obtained from \eqref{eq:00WWgam}, \eqref{eq:PPPCWW},
\eqref{eq:NLO00WWgam} and the unexpanded fixed-order result \cite{Bergstrom:2005ss}.
The shaded areas mark $1-x < 4
\epsilon$ and $x<4 \epsilon$. Note the plot is logarithmic 
in $1-x$.}
\label{fig:dNdx2TeV}
\end{figure}

In Fig.~\ref{fig:dNdx2TeV}, we show the $\gamma$-spectra obtained 
from~\eqref{eq:00WWgam}, \eqref{eq:PPPCWW}, and \eqref{eq:NLO00WWgam} for a wino of mass $\mchi = 3
\,{\rm TeV}$, together with the unexpanded (in $\epsilon$) fixed-order 
result \cite{Bergstrom:2005ss}. Close to the endpoint, here 
to the left side of the figure due to the logarithmic plot in 
$1-x$, the unexpanded fixed-order calculation and
the NLL'-accurate endpoint result match essentially perfectly 
inside the left-shaded band, which indicates the region of validity 
of the latter computation.\footnote{We emphasize that we compare fixed orders in the $\hat{\alpha}_2$ expansion. In the left-shaded band, 
higher-order corrections and large, and endpoint-resummation is 
essential. Similarly, in the right shaded band, $x<4\epsilon$, the fixed-order 
approximation is inaccurate.}  
The collinear approximation (PPPC4DM) 
fails badly here. In the transition region $1-x\;\lessim \;0.3$, 
fixed-order and the NLL' endpoint approximation still match well. 
The collinear approximation~\eqref{eq:PPPCWW} is about 20\% larger than the 
exact one in the region $0.1\;\lessim \;1-x\;\lessim\; 0.9$, where it should be merged 
to the endpoint-resummed approximation. In the final merged spectra 
presented below, this discrepancy in a subdominant channel is not a 
visible effect. 

To emphasize the structural dependence on the leading logarithms, we define $1-x = \beta \epsilon$, 
where $\beta$ is a fixed constant, such that we can expand all 
formulas for small $\epsilon$.
Doing so, yields for the leading $\epsilon^{-1}$ term
\begin{align}
\frac{dN_{WW}}{dx} &= \frac{2 \alpha_{\rm em}}{\pi} \,  \frac{1}{\epsilon} \left\lbrace \begin{array}{ccc}
\frac{\ln (1+\beta^2)}{\beta} - \frac{\beta}{1+\beta^2} & \quad \quad & {\rm int.\, res.} \\[0.1cm]
\frac{\ln \beta^2}{\beta}  & \quad \quad & {\rm PPPC4DM} \\[0.15cm]
\frac{\ln \beta^2 - 1}{\beta}  & \quad \quad & {\rm full\;fixed\;order}
\end{array} \right\rbrace + \mathcal{O}(\epsilon^0) \, .
\end{align}
As evident, for large $\beta$, the three approximations have the 
same leading behaviour, which is expected for the collinear approximation 
with respect to the full fixed-order computation. However, for the resummed endpoint 
result, this was not necessarily expected, as large $\beta$ violates 
the intermediate resolution scaling $1-x \sim \epsilon$ for which 
\eqref{eq:00WWgam} was derived. Furthermore, NLL' resummed and fixed order also yield the same
non-logarithmic term at large $\beta$, which is not part of the collinear
approximation, as it stems from model-dependent terms.

Turning to the charged wino annihilation process  
$\chi^+ \chi^- \to \gamma + X$, which is in fact the more 
important one (due to the Sommerfeld enhancement), the 
relevant final states are $X = W^+W^-, f \bar{f},
Zh$. In the collinear approximation, the $W^+W^-\gamma$ final 
state is produced by the same universal $W\to W\gamma$ splitting 
function, hence \eqref{eq:PPPCWW} applies for $\chi^+ \chi^-$ 
annihilation as well. On the other hand, extracting the 
$\chi^+ \chi^- \to W^+W^- \gamma$ process at lowest order 
from the resummed endpoint calculation gives
\begin{align}
\left. \frac{d N_{WW}}{dx}\right|^{{\rm int.}}_{\chi^+ \chi^- \to WW \gamma} &= \frac{2 \alpha_{\rm em}}{\pi} \left[- \frac{1-x}{\epsilon^2 + (1-x)^2} - \frac{3}{(1-x)} \ln \left(1 + \frac{(1-x)^2}{\epsilon^2}\right) \right. \nonumber \\
&\hspace{-2cm} \left.- \left(4 \ln \epsilon^2 + \frac{29}{8}\right) \left[\frac{1}{1-x}\right]_+ + 4 \left[\frac{\ln (1-x)}{1-x}\right]_+\right] \, .
\label{eq:pmWWgamma}
\end{align}
The plus-distributions regulate the limit $x \to 1$, and 
are taken care of together with the virtual contributions in the 
nominal zero-bin at the absolute
endpoint, discussed in Appendix~\ref{sec:zerobin}.
Since our ultimate goal is to provide a merged spectrum from the 
endpoint $x=1$ to $x \to 0$, we are particularly interested in the
region in between  $1-x \gg \epsilon$, where \eqref{eq:PPPCWW}
applies, and $1-x \sim \epsilon$, where  \eqref{eq:pmWWgamma} is 
valid. To this end, we expand \eqref{eq:pmWWgamma} for small $\epsilon$, and \eqref{eq:PPPCWW} in small $1-x$ to obtain 
\begin{align}
\left. \frac{d N_{WW}}{dx}\right|^{{\rm int.}, \,\epsilon\to 0}_{\chi^+ \chi^- \to WW \gamma} &= \frac{2 \alpha_{\rm em}}{\pi} \frac{1}{1-x} \left[ \ln \frac{1}{\epsilon^2} + \ln \frac{1}{(1-x)^2} - \frac{37 }{8} \right] + \mathcal{O}(\epsilon^2)\,, \\
\left. \frac{dN_{WW}}{dx}\right|^{1-x \to 0}_{\rm PPPC 4 DM} &=\frac{2 \alpha_{\rm em}}{\pi} \frac{1}{1-x} \left[ \ln \frac{1}{\epsilon^2} - \ln \frac{1}{(1-x)^2}  \right]   + \mathcal{O}((1-x)^0)\,.
\end{align}
As expected, the collinear approximation does not produce the correct 
logarithms as $x\to 1$, but the different sign between the 
$\ln (1-x)^{-2}$ terms ensures that the two
splitting functions \eqref{eq:PPPCWW} and \eqref{eq:pmWWgamma} 
intersect each other around $1-x \sim 0.3$. Furthermore, for large 
DM masses, i.e. small
$\epsilon$, the $\ln \epsilon^{-2}$ term ensures that the shape of 
both curves is similar in the region between $1-x \sim \epsilon$ 
and $1-x \gg \epsilon$. We employ a linear merging procedure 
between the two spectra, as detailed in Appendix~\ref{sec:merging}. 
Given the interplay of various terms due to the Sommerfeld 
enhancement factors, a full assessment of the merging
quality is possible only for the  full expressions for $dN/dx$ without 
expansion in $\hat{\alpha}_2$, 
see the following section.

The other two final states $\gamma f \bar{f},\gamma Z h$ contribute 
\begin{align}
\left. \frac{d (\sigma v)}{dx}\right|_{\chi^+ \chi^- \to \gamma f \bar{f},\gamma Z h} = (\sigma v)_{\chi^+ \chi^- \to \,2\gamma \gamma + \gamma Z } \, \frac{\hat{\alpha}_2}{\pi } \left[\frac{1}{1-x}\right]_+\left(1  + \frac{1}{48 \hat{c}_W^2}\right)\
\end{align}
to the NLL' endpoint-resummed spectrum. Here $(\sigma v)_{\chi^+ \chi^- \to \,2\gamma \gamma + \gamma Z } = 2 \pi
\hat{\alpha}_2 \hat{\alpha}_{\rm em}/\mchi^2$, and the expression 
includes the splitting into all SM fermion pairs and $Zh$, 
respectively, whose masses can be neglected at intermediate 
resolution. PPPC4DM does not include these processes. 

\section{Merged differential spectra}\label{sec:spectra}

\begin{figure}[t]
\centering
\includegraphics[width=0.78\textwidth]{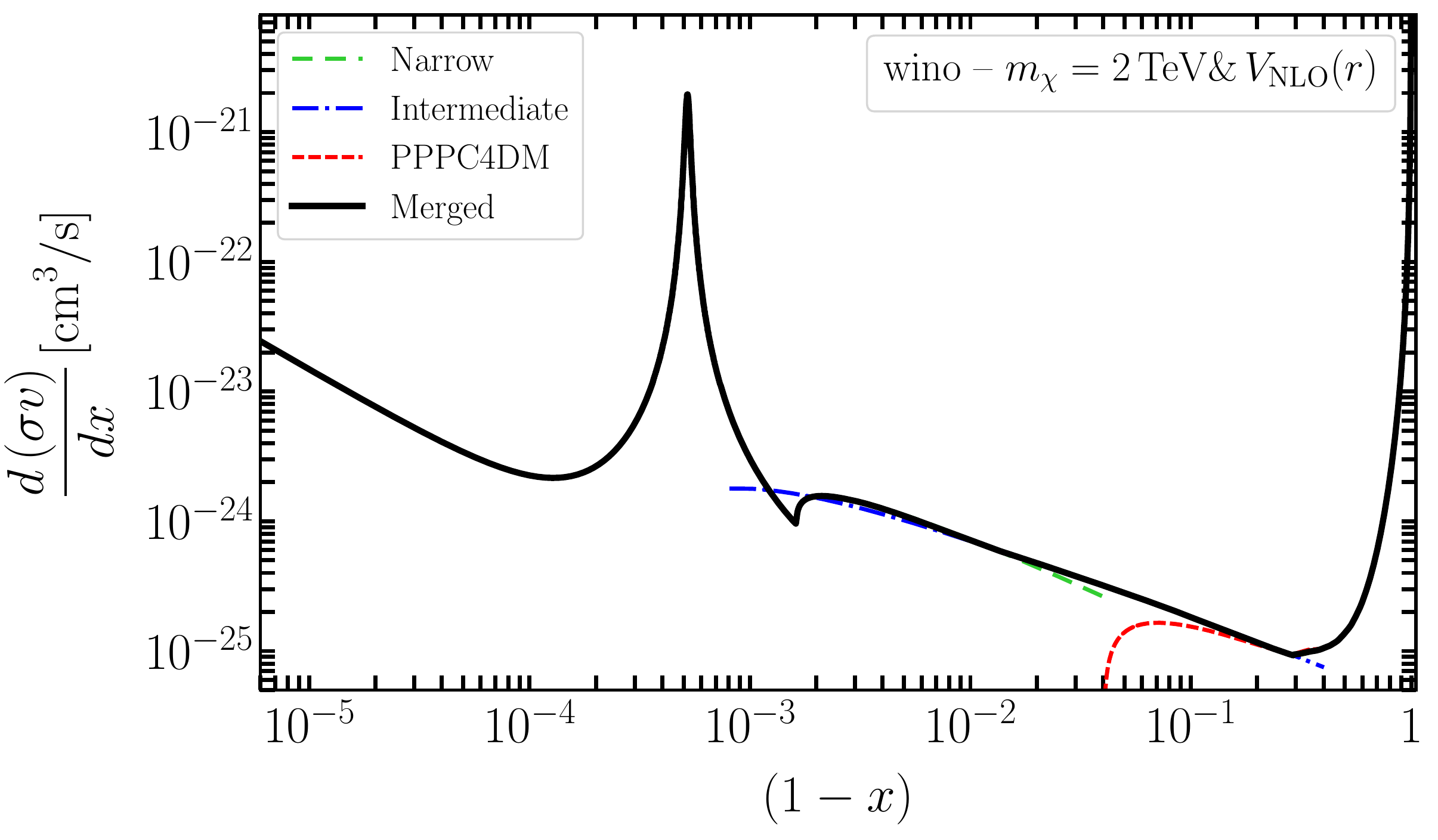}
\includegraphics[width=0.78\textwidth]{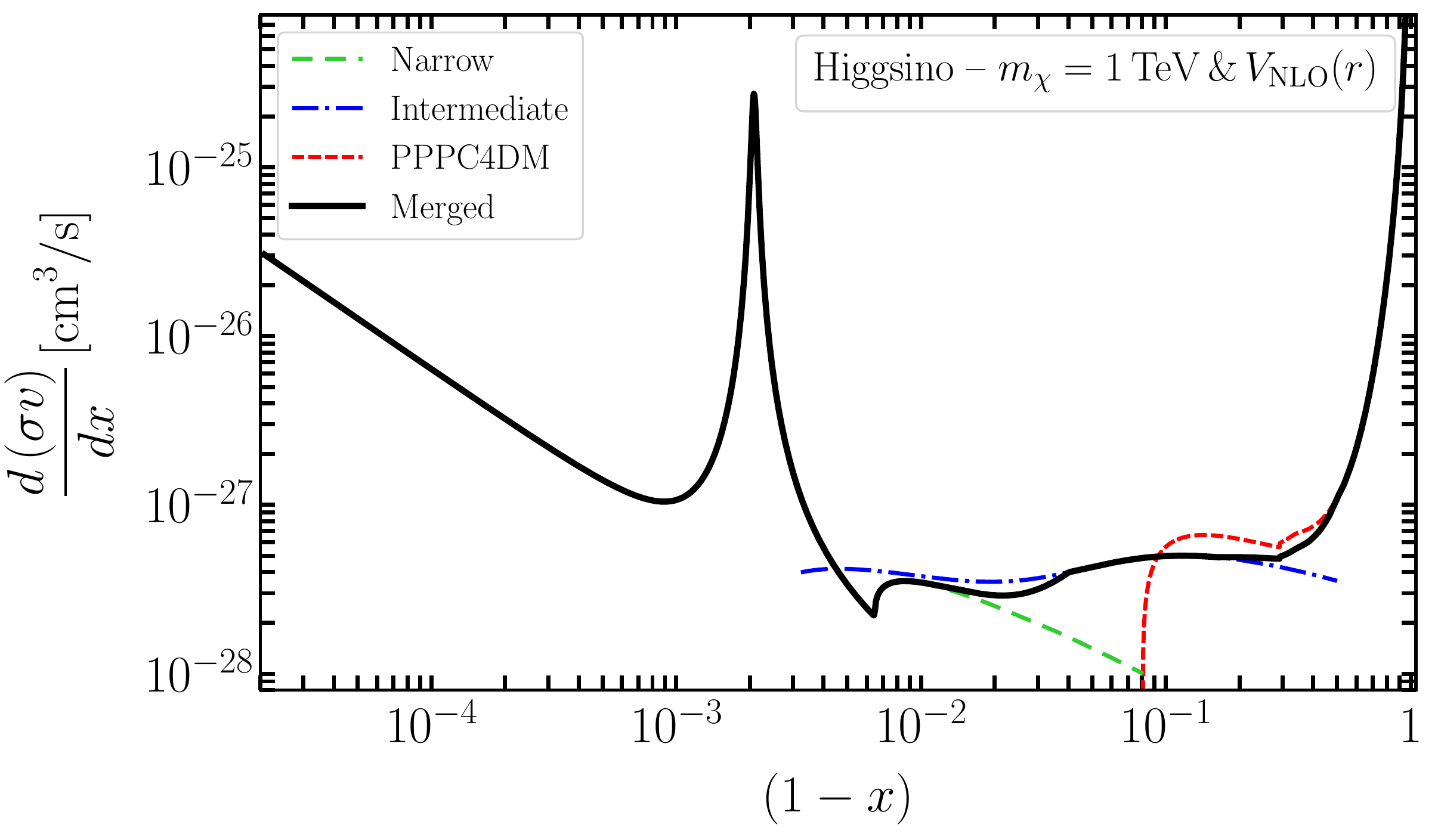}
\caption{Merged differential spectrum for a  $2\,{\rm TeV}$-mass wino (upper panel) and Higgsino with $\mchi = 1\,{\rm TeV}$ (lower panel) with NLO EW potentials and NLL' Sudakov resummation for the endpoint calculations. The merged spectrum is shown in black (solid), whilst the narrow and intermediate resolution calculations are depicted in green (dashed) and blue (dash-dotted), respectively. The PPPC4DM result (excluding the smeared $\delta$-distributions -- see Appendix~\ref{sec:PPPC4DMimpl}), is shown in red (short-dashed).
}
\label{fig:spectrum}
\end{figure}

To provide high-resolution fully differential spectra for 
all values of $x$, we merge 
the narrow resolution resummed calculation with the intermediate 
resolution resummation, and the latter with PPPC4DM. An in-depth 
investigation of the logarithmic structure of the narrow vs. 
intermediate resolution resummation to the two-loop order can be 
found in \cite{Beneke:2019gtg} and will not be repeated here. 
Technical details on the merging of the three results are
given in Appendix~\ref{sec:merging}.
In the following, we discuss the resulting photon energy spectrum 
for annhilation of a 2~TeV wino and a 1~TeV 
Higgsino, shown in Fig.~\ref{fig:spectrum}. The endpoint region
$1-x \to 0$ is depicted logarithmically to identify the opening 
of different final-state channels. 

Starting from the left, that is, at the endpoint of maximal photon 
energy, the spectrum to the left of the clearly visible 
$\gamma Z$ peak at  $1-x = \mZ^2/(4 \mchi^2)$ is caused by the process $\chi \chi \to \gamma f \bar{f}$, where 
$f$ are the SM fermions excluding the top quark. In the resummation
calculation, these fermions are taken as massless. 
The first light-fermion mass effects are expected from 
the bottom quark at $1-x = m_b^2/m_\chi^2$, which is already outside
the range shown in the figure. Light-fermion mass effects can be incorporated in a
straightforward modification of the narrow resolution result in
\cite{Beneke:2018ssm,Beneke:2019vhz,Beneke:2019gtg}. We refrain from
performing this modification, since the experimental capabilities
are by orders of magnitudes away from being able to 
resolve this effect. The width of the $\gamma Z$ peak arises from 
a consistent treatment of the $Z$-boson width using Dyson resummation
according to Eq.~(B.59) of \cite{Beneke:2019vhz}.\footnote{The analogous
expressions for hypercharge and mixed hypercharge/SU(2) narrow resolution
recoiling jet functions, which are not given in \cite{Beneke:2019vhz}, are
obtained by the simple replacement of the corresponding couplings and Weinberg angle.} To the right of the $Z$-peak, we observe a kink at $1-x = \mW^2/\mchi^2$
corresponding to the $\gamma W^+ W^-$ threshold opening up, with a collimated $W^+ W^-$ pair. The subsequent $Zh$ and $t \overline{t}$ 
thresholds lie already in the merging region between the narrow and intermediate resolution calculation, and would be too weak to be visible on the scale of the plot.

The regime of validity of the intermediate resolution
endpoint calculation begins around $1-x \sim \frac{\mW}{\mchi}$. 
Here, in addition to all the aforementioned processes, the emission of soft
$W$-radiation in all directions is kinematically possible. Also, 
soft initial-state radiation of electroweak gauge bosons 
becomes possible. At smaller $1-x$, soft 
effects are purely virtual for kinematic reasons. 
At $1-x\sim 0.3$, the intermediate resolution calculation is 
merged with the parton-shower calculation in PPPC4DM,\footnote{Note that PPPC4DM
is used in a modified form to avoid contamination with the smeared tree-level
$\delta$-distributions. For details, see Appendix~\ref{sec:PPPC4DMimpl}.} which replaces the dedicated endpoint calculations for the 
remaining part of the spectrum, which as expected, diverges for 
$x\to 0$, where the photon becomes soft, and the semi-inclusive 
photon spectrum is no longer observable.


We observe that for all practical purposes, there is an energy 
region where the two calculations to be merged are sufficiently 
accurate and in agreement, such that an accurate  spectral 
shape over the entire photon energy range from 0 to $m_\chi$ 
is obtained.

\begin{figure}[t] \centering
    \includegraphics[width=0.6\textwidth]{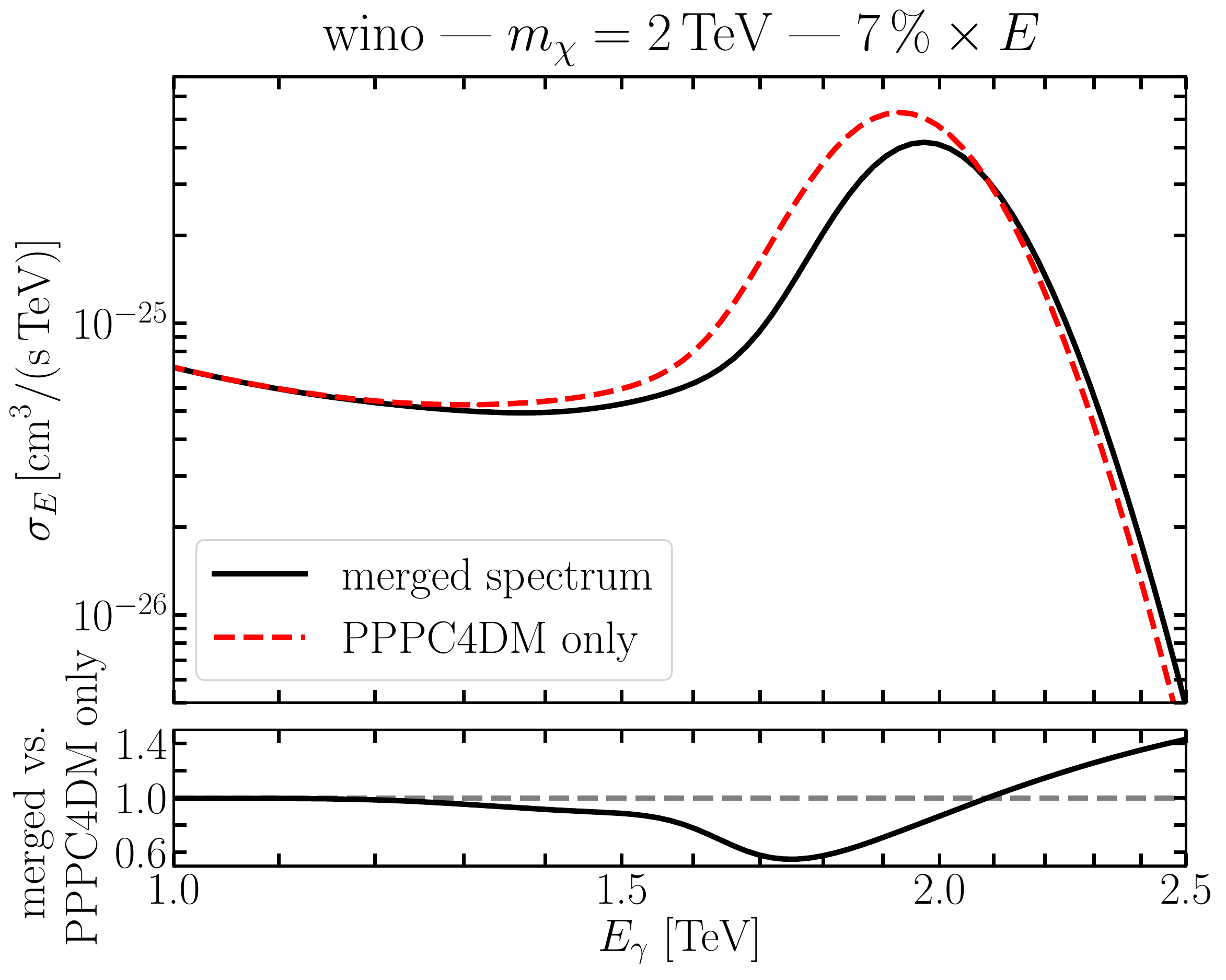}
    \includegraphics[width=0.6\textwidth]{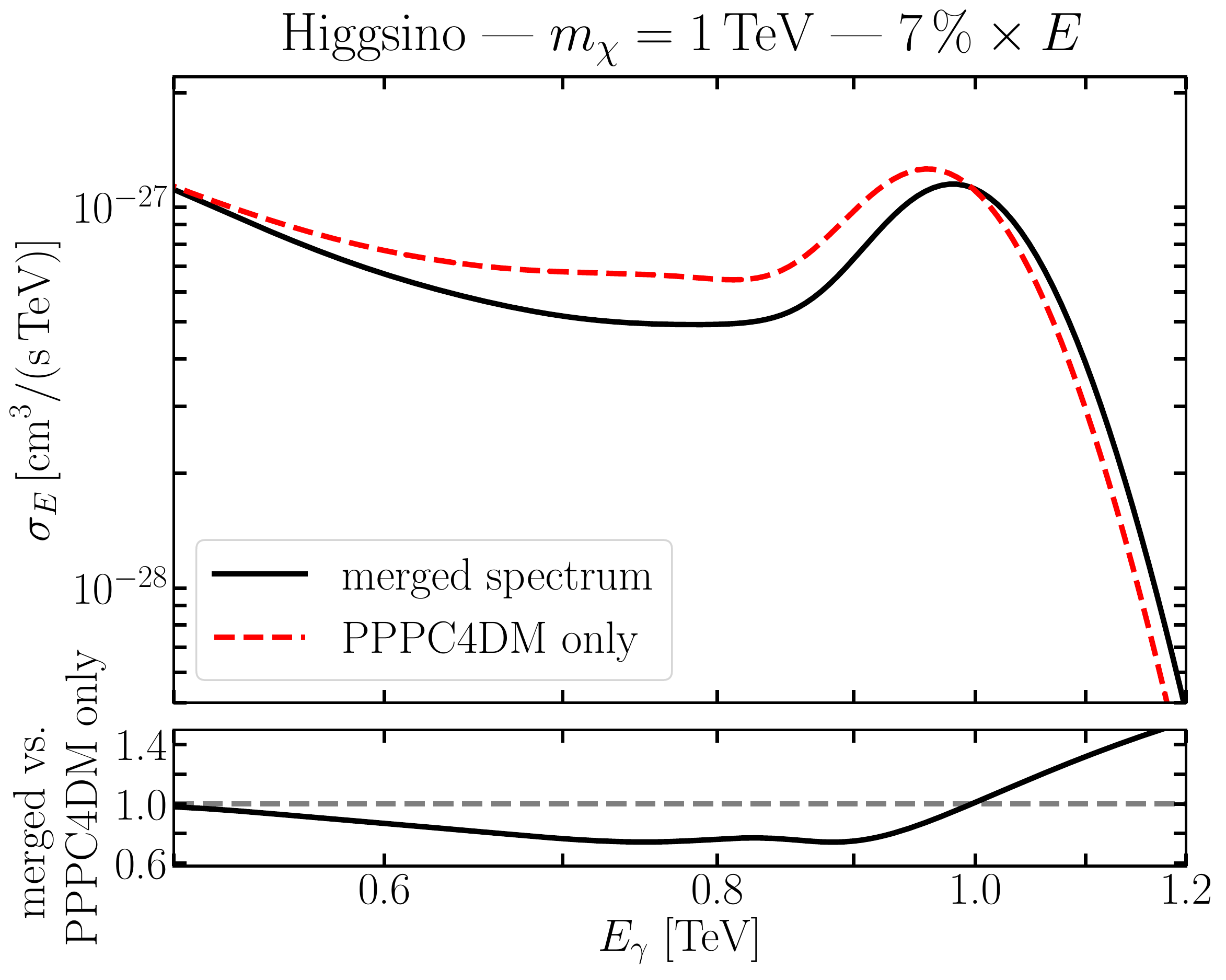}
    \caption{Mock folding with a Gaussian resolution of $7\% \times E_\gamma$ of the
        spectra of Fig.~\ref{fig:spectrum} for the $2\,{\rm TeV}$ wino (upper
        panel), and $1\, {\rm TeV}$ Higgsino (lower panel). The merged endpoint-        resummed spectrum is shown in black/solid, and the analogue PPPC4DM-only spectrum (including NLO Sommerfeld enhancement) in red/dashed.}
    \label{fig:gaussian} \end{figure}


\section{Discussion}
\label{sec:conclusion}

Is endpoint resummation important given the present and expected 
energy resolution of instruments? To address this question, we
fold the merged endpoint-accurate spectrum and the PPPC4DM-only 
spectrum with a Gaussian resolution function of energy resolution 
$a E_\gamma$,
\begin{align}
\sigma_E(a) \equiv \, \int_0^{m_\chi} d E_\gamma^\prime \,\,  \frac{d (\sigma v)}{d E_\gamma^\prime}\, \cdot \,\frac{1}{\sqrt{2 \pi}\cdot a \cdot E_\gamma} e^{-\frac{(E_\gamma - E_\gamma^\prime)^2}{2 \cdot a^2 \cdot E_\gamma^{2}}} \,,
\end{align}
and compare the two. The result is shown in Fig.~\ref{fig:gaussian} 
for the $2\,{\rm TeV}$ wino (upper plot) and $1\,{\rm TeV}$ 
Higgsino (lower plot), adopting an energy resolution of $7\,\% \times
E_\gamma$, as is expected for CTA in the TeV DM mass regime
\cite{Acharyya:2020sbj}. The merged spectrum (black/solid) and the 
PPPC4DM prediction omitting endpoint resummation (red/dashed) 
are shown in the upper panels of the plots. Far away from the 
endpoint $E_\gamma=m_\chi$, the two results agree by construction, 
since resummation is unimportant. In the endpoint region, however, 
large deviations can be seen even on the logarithmic scale of the 
plot, emphasizing the need for resummation. In the subtended panels, the ratio of the resummed, merged spectrum {vs.} the
PPPC4DM-only prediction is shown. For the merged spectrum of the wino, there is suppression 
of the signal relative to PPPC4DM till the endpoint and  
slightly beyond, until at some point the merged spectrum exceeds 
PPPC4DM. The reason for the latter is
that PPPC4DM smears the tree-level delta-distribution, whilst the EFT
computation does not need any smearing, hence the peak of the PPPC4DM-only
spectrum is pushed to smaller $E_\gamma$ than the nominal endpoint. 
The Higgsino spectrum exhibits the same features. In addition, near 
maximal photon energy, there is also an intricate
interplay between resummation and Sommerfeld enhancement, when 
the Higgsino mass is of order and below 1~TeV, that further adds 
to this enhancement of the resummed spectrum at and beyond the 
endpoint \cite{Beneke:2019gtg}. 

For higher DM masses than those shown in Fig.~\ref{fig:gaussian}, the 
differences between the PPPC4DM prediction and the merged, resummed 
spectra are even more pronounced, as the resummed EW logarithms grow, and the large Sudakov
suppression of the $\chi^+ \chi^-$ tree-level channel dominates 
the prediction. 

In summary, we find that electroweak resummation significantly 
changes the shape of the photon energy spectrum in the range 
$E_\gamma \sim (0.6\ldots 1)\, m_\chi$ and hence the form of the 
so-called ``line-signal''. The present work demonstrates on 
the example 
of the wino and Higgsino model that an accurate matching of 
endpoint-resummed calculations to the full energy spectrum can be 
performed.  

Ancillary to this paper, we provide the code \texttt{DM$\gamma$Spec} that produces the merged differential spectra shown in~Fig.~\ref{fig:spectrum} 
for DM masses in the range $(0.5 -100)\,{\rm TeV}$, together with other useful 
functions, such as cumulating the cross-section in energy bins. 
Furthermore, there is the option to use LO and NLO Sommerfeld calculations and different
Higgsino mass splittings. A short description of the functionality is given in
Appendix~\ref{sec:code}.

\subsubsection*{Acknowledgements}
We thank Torsten Bringmann, Stefan Lederer and Clara Peset for 
useful discussions. 
This work has been supported in part by the DFG 
Collaborative Research Center ``Neutrinos and Dark Matter in Astro- and 
Particle Physics'' (SFB 1258).

\appendix

\section{Technical details}
\label{app:technical}

\subsection{Merging procedures}
\label{sec:merging}
As the logarithms between narrow and intermediate resolution calculation match
accurately for an extended region of $1-x$ \cite{Beneke:2019vhz}, we merge the
narrow and intermediate differential spectra according to
\begin{align}
\left. \frac{d( \sigma v)}{d x} \right|_{\rm merged} = w_1(x,\varepsilon) \left. \frac{d (\sigma v)}{d x}\right|_{\rm narrow} + \left(1-w_1(x,\varepsilon)\right) \left. \frac{d (\sigma v)}{d x}\right|_{\rm intermediate}
\end{align}
with the simple linear function 
\begin{align}
w_1(x, \varepsilon) = \left\lbrace \begin{array}{llc}
0 &  \quad \mathrm{if} & \quad \varepsilon < 1-x \\
\frac{1}{1-4\varepsilon} \left( 1 - \frac{1-x}{\varepsilon}\right) &  \quad \mathrm{if} &\quad 4 \varepsilon^2 \leq 1-x \leq \varepsilon  \\[0.1cm]
1 &  \quad \mathrm{if} &\quad 1-x < 4 \varepsilon^2
\end{array}\right. \, .
\end{align}
The merging starts at $1-x  = 4
\epsilon^2$ in the center of the parametric validity region,  
and ends when the intermediate resolution calculation is fully
valid at $\epsilon = 1-x$.

Based on the findings of Sec.~\ref{sec:logs}, we devise a similar 
merging procedure between PPPC4DM and the intermediate calculation:
\begin{align}
\left. \frac{d( \sigma v)}{d x} \right|_{\rm full \, merged} = w_2(x,\varepsilon) \left. \frac{d (\sigma v)}{d x}\right|_{\rm merged} + \left(1-w_2(x,\varepsilon)\right) \left. \frac{d (\sigma v)}{d x}\right|_{\rm PPPC4DM} 
\end{align}
with
\begin{align}
w_2(x,\varepsilon) = \left\lbrace \begin{array}{lcc}
0 & \quad & \mathrm{if} \quad {\rm min}(20 \varepsilon,0.5)\leq 1-x \leq 1 \\
1- \frac{1-x-{\rm min}(4\varepsilon,0.2)}{{\rm min}(20 \varepsilon,0.5)-{\rm min}(4 \varepsilon,0.2)} & \quad & \mathrm{if} \quad {\rm min}(4\varepsilon,0.2) \leq 1-x \leq {\rm min}(20 \varepsilon,0.5)  \\
1 &  \quad & \mathrm{if} \quad 0 \leq 1-x \leq {\rm min}(4\varepsilon,0.2)
\end{array}\right. \, .
\end{align}
The boundaries for the start of merging the intermediate resolution and PPPC4DM
are located well in the intermediate
regime at ${\rm min}(4 \epsilon,0.2)$. The value $0.2$ is chosen, such that for small DM masses $\mchi
< 10 \mW$, the merging region is extended. Similarly, we assign the upper value
${\rm min}(20 \epsilon,0.5)$ to the merging region beyond which the spectrum is fully determined by
PPPC4DM.  

Note that there is no overlap between the two merging steps for the DM masses
relevant in this paper, as ${\rm min}(4 \epsilon,0.2) > \epsilon$, as long as
$\mchi > 2.5 \mW$, i.e. for DM masses $\mchi > 200 \,{\rm GeV}$.

\subsection{Zero-bin}
\label{sec:zerobin}

In representing the spectra in the main text, we focused on the differential
spectrum and excluded the absolute endpoint $x=1$ from the figures. The endpoint
spectra are distribution-valued objects requiring careful treatment at the absolute endpoint. For example, the differential spectrum
contains delta-distributions associated with the tree-level $\chi \chi \to
\gamma \gamma$ process and virtual corrections to this process, which are
obviously not accessible in a plot of the result. In addition, the result
depends on plus-distributions that arise from the expansion of
\begin{align}
\frac{1}{(1-x)^{1-\eta}} = \frac{\delta(1-x)}{\eta} + \sum_{n=0}^\infty \frac{\eta^n}{n!} \left[\frac{\ln^n (1-x)}{1-x}\right]_+ \,.
\end{align}
These contain subtractions if integrated to the endpoint. For example, for the
ordinary plus-distribution
\begin{align}
\int_0^1 d(1-x) f(1-x) \cdot \left[\frac{1}{1-x}\right]_+ =  \int_0^1 d(1-x) \frac{f(1-x) - f(0)}{1-x} \,.
\end{align}
To make the spectra easily accessible without having to worry about the
implementation of the distributions, we provide in the code a zero-bin, which is the integral of the spectrum including $\delta(1-x)$ terms from $\chi \chi \to \gamma \gamma$
from the absolute endpoint $x = 1$ to $x= 1 - 2 \cdot 10^{-5}$. The zero-bin takes care of all the distributions and their integration.
\subsection{Implementation of PPPC4DM}
\label{sec:PPPC4DMimpl}

When using PPPC4DM \cite{Cirelli:2010xx} to merge with the endpoint-resummed
spectra, we perform some modifications to obtain a consistent result. We
exclude the tree-level final states $\gamma \gamma,\gamma Z,ZZ$ in
\eqref{eq:dNdx_def} (and therefore also their subsequent splittings) 
between the endpoint and $1-x = 1- 10^{-0.15}$ where effects of
the otherwise smeared version of tree-level delta-distributions are visible. These
tree-level pieces are part of the endpoint calculation, and hence to avoid
double-counting need to be removed from PPPC4DM.
We confirmed analytically and numerically, following \cite{Ciafaloni:2010ti}, that in this
modified region, the neglected contributions are numerically subleading to the
splitting $dN_{WW}/dx$.

In addition, we extend the interpolating tables provided by \cite{Cirelli:2010xx} for
$dN_{WW}/dx$ in the modified region discussed above, since the interpolating table is
too sparse in the absolute endpoint region. To this end, we use the pure Monte
Carlo data of \cite{Cirelli:2010xx} and apply the EW evolution following
\cite{Ciafaloni:2010ti}. We confirm the few points provided by PPPC4DM in this
region and, by adding additional points to the interpolation, obtain a smooth
interpolating function that is easily merged into the endpoint-resummed
results.

\section{Spectrum code -- \texttt{DM$\gamma$Spec}}
\label{sec:code}

Ancillary to this paper, we provide the code package \texttt{DM$\gamma$Spec} that produces differential
spectra (endpoint resummed to NLL' and merged to PPPC4DM as discussed in the
main text) and some other functions for the wino and Higgsino model in the DM mass range $\mchi = (0.5-100)\,{\rm TeV}$. Here we summarize the main functionality. A detailed account,
including the code validation tests, is provided with the code in separate
documentation at 
\begin{align*}
\href{https://dmyspec.hepforge.org}{\mathtt{dmyspec.hepforge.org}}
\end{align*}
or alternatively from \href{https://users.ph.tum.de/t31software/dmyspec}{\texttt{https://users.ph.tum.de/t31software/dmyspec}}. The code requires a \texttt{Python 3} installation and the packages
\texttt{numpy} and \texttt{scipy}. The main functions described below can be
loaded with
\begin{align*}
\mathtt{from\ \ resummation\ \  import\ \ function\_name}
\end{align*}
in a \texttt{Python} interpreter. Another (recommended) possibility is to use the provided example Jupyter
notebook file (requires Jupyter notebooks installed), that can be loaded, e.g.,
via 
\begin{align*}
\mathtt{jupyter\ \,  notebook\ \, ./example\_data.ipynb}
\end{align*}
in a Unix Shell. In the following, we describe the main functions of the code.
An in-depth discussion including validation and further installation 
information can be found in the accompanying documentation on the webpage.

\subsection{diffxsection}
Provides the differential cross-section $d (\sigma v) / dx$ in $x$ for $\chi^0 \chi^0 \to \gamma + X$ for all
values of~$x$, except for $1-x \leq 2 \cdot 10^{-5}$, where the zero-bin at the
absolute endpoint (see below) has to be used. The output of 
\begin{align}
    \mathtt{diffxsection(x,mchi,model,SF)}
\end{align}
 has dimension
$10^{-26} {\rm cm^3}/{\rm s}$. The function parameters refer to the variable $\mathtt{x} = E_\gamma /
m_\chi$, and $\mathtt{mchi}$ to the DM mass in units of TeV. The parameter
$\mathtt{model}$ specifies if the Higgsino or wino model shall be investigated,
possible values are \texttt{\textquotesingle wino\textquotesingle} or
\texttt{\textquotesingle higgsino\textquotesingle} (supplied as a Python
string, i.e., including the quotation marks).  

Finally there is the parameter
\texttt{SF} that specifies the Sommerfeld-factor table to be used. The Sommerfeld
factor, apart from the EW potential and the mass difference between partners in the
multiplet, also depends on the velocity 
of the lightest DM particle, denoted by
$v$. For the model \texttt{\textquotesingle wino\textquotesingle}, 
the identifier \texttt{SF} can take values from the list 
\vskip0.2cm\noindent
{\small \centerline{
\texttt{\textquotesingle LO -3\textquotesingle},
\texttt{\textquotesingle LO -4\textquotesingle},
\texttt{\textquotesingle LO -5\textquotesingle},
\texttt{\textquotesingle NLO -3\textquotesingle},
\texttt{\textquotesingle NLO -4\textquotesingle},
\texttt{\textquotesingle NLO -5\textquotesingle},}}
\vskip0.2cm\noindent
where either the  \texttt{LO} or 
\texttt{NLO} Sommerfeld potential is used, and 
the negative integer $n$ refers to the exponent of the 
(single-particle) velocity $v=10^n$, at which the Sommerfeld enhancement is evaluated. 
The mass splitting between the charged and neutral wino is 
always fixed to  $\delta m_\chi=164.1\,$MeV.
For the model \texttt{\textquotesingle higgsino\textquotesingle}, 
the identifier \texttt{SF} can take the following 12 values:
\vskip0.2cm\noindent
{\small \hskip0.5cm\texttt{\textquotesingle LO -3 dm 355 dmN 20\textquotesingle},\; 
\texttt{\textquotesingle LO -4 dm 355 dmN 20\textquotesingle},\;
\texttt{\textquotesingle LO -5 dm 355 dmN 20\textquotesingle},\\
\hspace*{0.5cm}\texttt{\textquotesingle NLO -3 dm 355 dmN 20\textquotesingle},\;
\texttt{\textquotesingle NLO -4 dm 355 dmN 20\textquotesingle},\;
\texttt{\textquotesingle NLO -5 dm 355 dmN 20\textquotesingle},\\
\hspace*{0.5cm}\texttt{\textquotesingle LO -3 dm 355 dmN 015\textquotesingle},\;
\texttt{\textquotesingle LO -4 dm 355 dmN 015\textquotesingle},\;
\texttt{\textquotesingle LO -5 dm 355 dmN 015\textquotesingle},\\
\hspace*{0.5cm}\texttt{\textquotesingle NLO -3 dm 355 dmN 015\textquotesingle},\;
\texttt{\textquotesingle NLO -4 dm 355 dmN 015\textquotesingle},\;
\texttt{\textquotesingle NLO -5 dm 355 dmN 015\textquotesingle}}
\vskip0.2cm\noindent
The first two entries have the same intepretation as for the wino 
model. The number following \texttt{dm} specify the mass splitting 
between $\chi_1^0$ and $\chi^+$, which always takes the value 
$\delta m_\chi=355\,$MeV. The last number specifies the mass 
difference between the two neutral Higgsinos $\chi_1^0$ and 
$\chi_2^0$, and we provide tables for the two cases $\delta m_N = 
20\,$MeV and $150\,$keV.

\subsection{cumulxsection}

This function cumulates the cross section from the endpoint $x=1$ to a given $x
\leq 1$, i.e.,
\begin{align}
    \int^1_{1-x} dx^\prime \, \frac{d (\sigma v)}{dx^\prime} \, .
\end{align}
Note that if $1-x$ falls within the zero-bin, the integration is extended to
the zero-bin size (see below). The output of
\begin{align}
    \mathtt{cumulxsection(x,mchi,model,SF, ZBsize = \text{\texttt{\textquotesingle default\textquotesingle}}, rel = -3)}
\end{align}
 has dimension $10^{-26} {\rm cm^3} /
{\rm s}$ with function arguments as for \texttt{diffcross} above. In addition, there is
the optional function value \texttt{ZBsize} that if omitted is set to
\texttt{\textquotesingle default\textquotesingle}. The possible parameters for
this option are either \texttt{\textquotesingle default\textquotesingle}
corresponding to a zero-bin of width $1-x=2 \cdot 10^{-5}$ or
\texttt{\textquotesingle 1 \%\textquotesingle} for a zero-bin width of $1-x = 0.01$.
The parameter \texttt{rel} refers to the relative error requirement for the
integrator and by default is set to $\mathtt{-3}$ corresponding to a relative
error of $10^{-3}$ (for an in-depth discussion, see the accompanying documentation).

\subsection{binnedcross}

Similar function to \texttt{cumulcross} above, however, for a chosen energy bin
from $E_1$ to $E_2$ with $E_1 < E_2$, i.e.,
\begin{align}
    \int^{E_2/m_\chi}_{E_1/m_\chi}dx \frac{d (\sigma v)}{dx}\, .
\end{align}
The output of 
\begin{align}
    \mathtt{binnedxsection(mchi,E1,E2,model,SF,ZBsize = \text{\texttt{\textquotesingle default\textquotesingle}}, rel = -3)}
\end{align}
has dimension $10^{-26} {\rm cm^3}/{\rm s}$, and the function parameters are as for \texttt{cumulcross} above, with the
addition of $E_1 < E_2 \leq m_\chi$ both given in units of TeV instead of
\texttt{x}.

\subsection{zerobin}

To allow for the inclusion of the absolute endpoint, at 
which the exclusive $\chi \chi \to \gamma \gamma$ channel 
is located, a zero-bin has to be provided
(cf.~\ref{sec:zerobin}).  The zero-bin ranges are $1-x = 0$ to $1-x=2 \cdot
10^{-5}$ (\texttt{\textquotesingle default\textquotesingle}) or to
$0.01$ (\texttt{\textquotesingle 1 \%\textquotesingle}).  The 
output of 
\begin{align}
    \mathtt{zerobin(mchi,model,SF,ZBsize = \text{\texttt{\textquotesingle default\textquotesingle}})}
\end{align}
has dimension $10^{-26} {\rm cm^3}/{\rm s}$. The function
arguments are analogous to \texttt{cumulcross} above.

\bibliography{dm}

\end{document}